\documentclass[reprint,aps,amsmath,amssymb,nofootinbib,twoside,prd,showkeys,superscriptaddress,floatfix]{revtex4-2}
\pdfoutput=1
\usepackage{verbatim}
\usepackage{graphicx}
\usepackage[T1]{fontenc}
\usepackage{epsfig}
\usepackage{bm}
\usepackage{tensor}
\usepackage{amssymb}
\usepackage{amsmath}
\usepackage{subfigure}
\usepackage{dcolumn}
\usepackage[utf8]{inputenc}
\usepackage{cancel}
\usepackage[usenames,dvipsnames]{color}
\usepackage[colorlinks]{hyperref}
\hypersetup{
     breaklinks=true,
    pdfstartview={FitH},    
    colorlinks=true,       
    linkcolor=blue,          
    citecolor=red,        
    filecolor=magenta,      
    urlcolor=blue,           
    anchorcolor=green,      
    linktocpage=true
}

\def\doi{http://doi.org}

\newcommand{\eref}[1]{Eq.~(\ref{#1})}

\newcommand{\mnras}{Mon. Not. R. Astron. Soc.}   

\newcommand{\HCd}{\mathcal{H}}

\def\HCdt0{\tilde{\HCd}_{0}}

\newcommand{\rs}{r_\mathrm{s}}

\usepackage{multirow}
\usepackage{amssymb}	
\usepackage{graphicx}
\usepackage{tikz}

\newcommand{\afffias}{Frankfurt Institute for Advanced Studies (FIAS), Ruth-Moufang-Strasse~1, 60438 Frankfurt am Main, Germany}

\newcommand{\affjwg}{Fachbereich Physik, Goethe-Universit\"at, Max-von-Laue-Strasse~1, 60438~Frankfurt am Main, Germany}
\newcommand{\affgsi}{GSI Helmholtzzentrum f\"ur Schwerionenforschung GmbH, Planckstrasse~1, 64291 Darmstadt, Germany}
\newcommand{\affHIP}{Helsinki Institute of Physics, P.O. Box 64, FI-00014 University of Helsinki, Finland}

\begin{document}
\title{ {Bounding the Cosmological Constant using Galactic Rotation Curves from the SPARC Dataset}}
\author{David Benisty}
\email{benidav@post.bgu.ac.il}
\affiliation{\afffias}\affiliation{\affHIP}
\author{David Vasak}
\email{vasak@fias.uni-frankfurt.de}
\affiliation{\afffias}
\author{J\"urgen Struckmeier}
\email{struckmeier@fias.uni-frankfurt.de}
\affiliation{\afffias}
\author{Horst Stoecker}
\email{stoecker@fias.uni-frankfurt.de}
\affiliation{\afffias}\affiliation{\affjwg}\affiliation{\affgsi}
\begin{abstract}
Dark energy (and its simplest model, the Cosmological Constant or $\Lambda$) acts as a repulsive force that opposes gravitational attraction. Assuming galaxies maintain a steady state over extended periods, the estimated upper limit on $\Lambda$ studies its  {pushback} to the attractive gravitational force of dark matter.  {From the SPARC dataset, we select galaxies that are best fitted by the Navarro-Frenk-White (NFW) and Hernquist density models.} Introducing the presence of $\Lambda$ in these galaxies helps  {to} establish the upper limit on its repulsive force. This upper limit on $\Lambda$ is around $\rho_{\left(<\Lambda\right)} \sim 10^{-25}$~kg/m$^3$, only two orders of magnitude higher than the one measured by Planck.  {We show that for galaxies with detectable velocities far from the galaxy core, the upper limit on $\Lambda$ is lower. Furthermore, we show that galaxies and other systems follow the same principle: for larger orbital periods the upper limit on $\Lambda$ is lower.} Consequently, we address the implications for future measurements on the upper limit and the condition for detecting the impact of $\Lambda$ on galactic scales. 
\end{abstract}
\maketitle
\section{Introduction}
\label{sec:Introduction}
DARK ENERGY (DE) is an unknown form of energy that affects the universe on the largest scales~\cite{Peebles:2002gy}. Its primary effect is to drive the accelerating expansion of the universe.  {In Einstiens' general relativity it is represented by a simple model, the cosmological constant $\Lambda$ at the focus in this paper. The first evidence for $\Lambda$ existence came from observations of supernovae~\cite{SupernovaCosmologyProject:1998vns,Pan-STARRS1:2017jku}. In addition to this, the Baryon Acoustic Oscillations (BAO)~\cite{Addison:2013haa,Aubourg:2014yra,Cuesta:2014asa,Cuceu:2019for} and the Cosmic Microwave Background (CMB)~\cite{Planck:2018vyg} give strong evidence for the need to modify either the matter sector of the universe or the gravitational sector.}

 {Dark matter is the second dominant part in our Universe as its density is about five times larger than the baryon matter density. Its postulation is motivated by observations of the dynamics of galaxies and their flat rotation curves~\cite{Trimble:1987ee}.} Without dark matter the slope of the rotation curve, which is the plot of the orbital velocities of visible stars or gas in a galaxy vs their radial distance from the center, would be much steeper.

From the observation of different galaxies like Andromeda, the Milky Way~\cite{Calcino:2018mwh,Gomes-Oliveira:2023qcr,Zhang:2023neo}, dwarf galaxies and others~\cite{Lelli:2016zqa,Lelli:2016uea,McGaugh:2016leg,2016ApJ...816L..14L,2017ApJ...836..152L,2017MNRAS.466.1648K,Lelli:2017sul,Desmond:2018oai,Li:2018rnd,Li:2018tdo,Katz:2018wao,2018NatAs...2..924M,2018MNRAS.480.2292S,2018MNRAS.480.4287K,Li:2019zvm,McGaugh:2019auu,Lelli:2019igz,2019MNRAS.483.1496S,Street:2022nib}, a universal density profile, now known as the Navarro-Frenk-White (NFW) model that grows at small radii like $\rho \sim r^{-1}$ and behaves as $\rho \sim r^{-3}$ for large radii, was first suggested in Ref.~\cite{Navarro:1996gj}. Subsequent work, though, weakened the universality of that model and proposed alternative density profiles~\cite{Sp08,Na10,Zhao:1995cp,Ev05,Ev14,Lin:2019yux,Hayashi:2020jze}.

Modified Gravity (MoG) can mimic the dark matter components~\cite{Benisty:2021sul,vandeVenn:2022gvl,Kirsch:2023iwd}. For example the Modified Newtonian Dynamics (MOND) model was explicitly designed to explain the flatness of rotation curves~\cite{Milgrom:1983ca,Bekenstein:1984tv,Kroupa:2018kgv}. It modifies Newton's second law at low accelerations and recovers the Tully-Fisher relation without introducing additional dark matter~\cite{Tully:1977fu,McGaugh:2000sr,Chae:2020omu,Bekenstein:2004ne,Piazza:2003ri,Khelashvili:2024gus}. Another ansatz for modified gravity is to define a different potential at the low energy limit by integrating over the dark matter model~\cite{Capozziello:2017rvz,Naik:2018mtx,Naik:2019moz,deAlmeida:2018kwq,Craciun:2023bmu}. Tests of MoG have been carried out using a variety of probes from the kinematics of stars and gas in the Milky Way~\cite{Henrichs:2020cae} through the fundamental plane of elliptical galaxies~\cite{Capozziello:2020dvd} to galaxy rotation curves~\cite{Hernandez-Arboleda:2022rim,Mannheim:2010xw,Oppenheim:2024rcp}.

 {Although $\Lambda$ dominates on cosmological scales, the expansion of space is also effective on the scale of the local universe~\cite{Chernin:2000pq,Baryshev:2000kw,Chernin:2001nu,Kim:2020gai,Karachentsev:2003eh,Chernin:2003qd,Teerikorpi:2005zh,Chernin:2009ms,Chernin:2006dy,Peirani:2008qs,Chernin:2010jt,Teerikorpi:2010zz,Chernin:2015nna,Chernin:2015nga,Silbergleit:2019oyx,Nandra:2011ui,Benisty:2023vbz,Benisty:2024lsz}.} In the low energy regime, the Newtonian force acting on a mass $M$ is modified by the repulsion force from $\Lambda$:
\begin{equation}
\frac{\ddot{r}}{r} = - G\,\frac{M}{r^3} + \frac{\Lambda c^2}{3}.
\label{eq:force}
\end{equation}
Here $r$ is the distance of the star to the mass center, $G$ is the Newtonian gravitational constant, and $c$ is the speed of light. As shown in Ref.~\cite{Benisty:2023vbz,Benisty:2023clf}, the impact of dark energy can indeed be observed also in the local Universe when the Milky Way and the M31 galaxy are considered as a binary system. In this paper, we wish to assess for the first time the influence of dark energy on galaxy rotation curves, and estimate an upper limit on the cosmological constant. We show that albeit $\Lambda$ is naturally related to a length scale, there is a strong correlation between the upper limit of $\Lambda$ and the galaxy scaling density.

As Ref.~\cite{Benisty:2023clf} shows, for the shorter periods of the system, the upper limit on $\Lambda$ will be reduced. This work extends Ref.~\cite{Benisty:2023clf} from binary motion to a spherical density in general, and galaxies in particular. Moreover, while Ref.~\cite{Benisty:2023clf} correlates the frequency of the binary motion to the upper limit on $\Lambda$, this work correlates curvature with the scaling density of the galaxy.

The structure of the paper is as follows: Section~\ref{sec:thoery} develops the theoretical framework for a spherical density model with $\Lambda$. Section~\ref{sec:method} describes the data set and the method for constraining $\Lambda$ by an upper limit. Section~\ref{sec:probes} compares the results with other systems. Section~\ref{sec:dis} provides an outlook on further research and discusses the efficiency of bounding $\Lambda$.

\section{Flat Rotation Curves with Dark Energy}
\label{sec:thoery}
\subsection{The Spacetime}
The line element $ds^2$ for the de~Sitter-Schwarzschild spacetime reads in spherical coordinates:
\begin{align}
&ds^2 = - \left(1 - 2\Phi(r)/c^2\right) dt^2 + \frac{dr^2}{1 - 2\Phi(r)/c^2} \nonumber
\\ & +  r^2 \left[\sin^2\theta \, d\theta^2 +  d\phi^2\right] \;,  
\label{eq:le}
\end{align}
with the potential: 
\begin{equation}
\Phi(r) = \frac{G M(r)}{r} + \frac{1}{6}\Lambda c^2 r^2,
\label{eq:f}
\end{equation}
where $M(r)$ is the integrated density up to the radius~$r$. The potential describes the gravitational and dark-energy contributions.

It is thus a de~Sitter-Schwarzschild metric which reduces to the Schwarzschild metric in the limit of $\Lambda=0$. For tracking the impact of the cosmological constant in different systems, we use the Kretschmann scalar~\cite{Cherubini:2002gen}:
\begin{equation}
\mathcal{K} \equiv R^{\lambda}{}_{\alpha\beta\gamma} R_{\lambda}{}^{\alpha\beta\gamma} \;,
\end{equation}
in which $R^{\lambda}{}_{\alpha\beta\gamma}$ is the Riemann curvature tensor. For a spherical density model in the de~Sitter-Schwarzschild metric, the Kretschmann scalar at the distance $r$ from the center reads: 
\begin{equation}
\mathcal{K} = 48 \left(\frac{G M(r)}{c^2 r^3} \right)^2+ \frac{2}{3} \Lambda^2 \;.  
\label{eq:K}
\end{equation}

\begin{figure}[t!]
\centering
\includegraphics[width=0.45\textwidth]{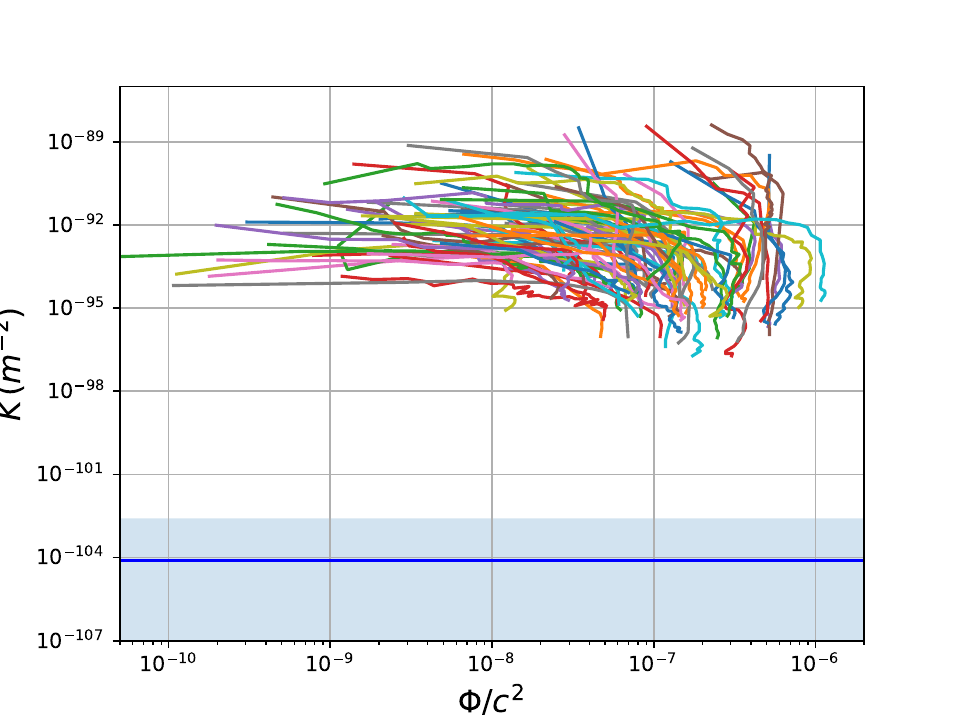}
\caption{\it{The normalized potential $\Phi/c^2$ vs.\ the scalar-curvature $K$ of galaxies extracted directly from the SPARC dataset via the measured circular velocities. The galaxy curves are parameterized by the distance from its center.  {The Planck value of the cosmological constant is the blue line.} }}
\label{fig:phase}  
\end{figure}

The observational data for the rotation curves considered in this work are taken from the catalog Spitzer Photometry \& Accurate Rotation Curves \texttt{(SPARC)}\footnote{\url{http://astroweb.cwru.edu/SPARC/}}sample~\cite{Lelli:2016zqa}. As a rough estimation for the potential and the curvature, we use combinations with $v(r)$ and $r$ (assuming $v_i^2 = G M(r_i)/r_i$): 
\begin{equation}
\frac{\Phi}{c^2} = \left( \frac{ v_i }{c} \right)^2 + \frac{1}{6} \Lambda r_i^2, \quad 
\mathcal{K} = 48 \left(\frac{v_i}{c r_i}\right)^4 + \frac{2}{3} \Lambda^2,
\end{equation}
where $v_i = v(r_i)$ and $r_i$ are the individual velocities and radii from the data. Figure~\ref{fig:phase} shows the potential vs.\ the curvature scalar for the SPARC dataset. We see that there is a few orders difference between the standard curvature of the galaxy and the curvature from the cosmological constant (solid line in blue), which indicates that we may only get an upper limit, but not a detection for~$\Lambda$. However, it seems that with better data we may reduce the upper limit and, as we shall see, push the upper limit closer to the actual value. 

\subsection{Spherical Dark-Matter Models} 
The modeling of dark halos is traditionally done by assuming a density profile. For two-power density models, the luminosity density of many elliptical galaxies can be approximated by a power law in radius at both the largest and smallest observable radii, with a smooth transition between these power laws at intermediate radii. Numerical simulations of the clustering of dark-matter particles suggest that the mass density within a dark halo has a similar structure. We test two different models, the NFW and the Hernquist model~\cite{1990ApJ...356..359H}:
\begin{equation}
\bar{\rho} (x) :=  \frac{\rho(x)}{\rho_s} =
\begin{cases} \left(x \left( 1 + x \right)^2\right)^{-1} & \text{NFW},\\
\left(x \left( 1 + x \right)^3\right)^{-1}  & \text{Hernquist}, \end{cases}
\end{equation}
 {where $x = r/r_s$ is the normalized radius relative to $r_s$. $r_s$ is scaling radius, where the dark matter density profile changes its behavior. For the NFW profile, for example, the density behaves as $\rho \sim r^{-1}$ for $r \gg r_s$, and as $\rho \sim r^{-3}$ for $r \ll r_s$. $r_s$ sets the scale for the density profile and is therefore a characteristic length parameter for the size of the galaxy ($r_{200}$). Relating the analysis to $r_s$ makes different galaxies comparable. $\rho_s $ is the average density within the scaling radius: 
\begin{equation}
  \rho_s := M(r_s)/\left(4\pi\,r_s^3/3\right).  
\end{equation}
} 
The corresponding mass density for these models gives the following:
\begin{equation}
\bar{M} (x) = \frac{M(x)}{16 \pi \rho_s r_s^3} =  \begin{cases} \ln(1 + x) - \frac{x}{1+x} & \text{NFW},\\\frac{x^2}{2 \left(1+x^2\right)}  & \text{Hernquist}. \end{cases}
\end{equation}

\begin{figure}[t!]
\centering\includegraphics[width=0.49\textwidth]{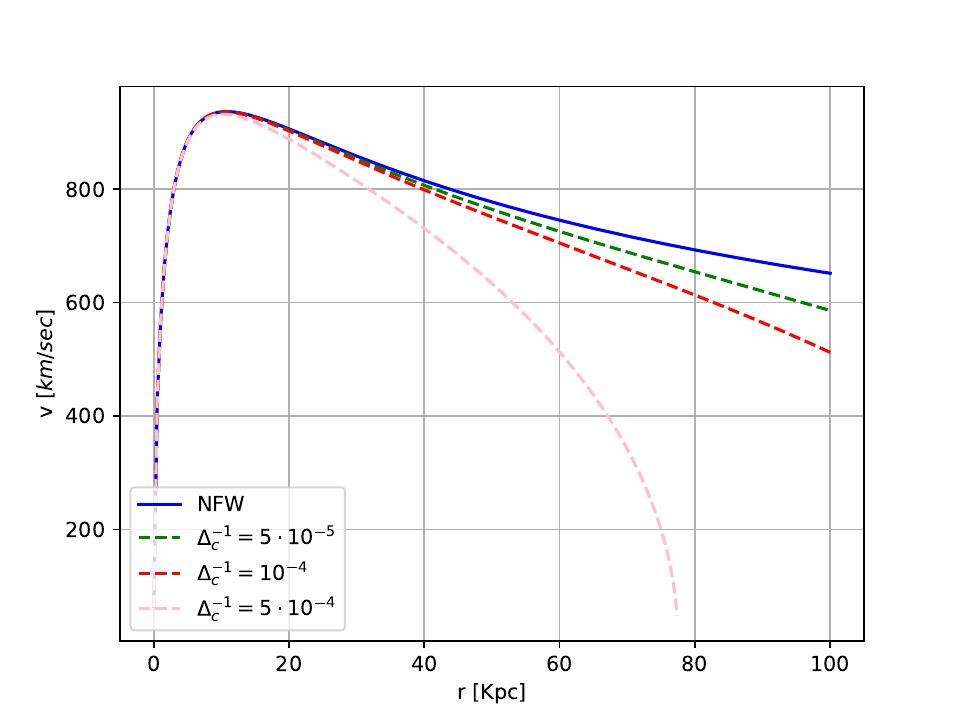} 
\caption{\it{The circular velocity of galaxy for the NFW profile with and without different values of DE. For larger $\Lambda$, the maximal velocity is reduced and sets the boundary for the galaxy flat rotation curves.}}
\label{fig:NFWvel} 
\end{figure}

\subsection{Halo velocity in the presence of \texorpdfstring{$\Lambda$}{Lambda}} 
The equations of motion of a massive particle in the gravitational field described by the general spherically symmetric metric can be obtained from the geodetic equation. Ref.~\cite{Nandra:2011ui} shows that the circular velocity $v_\mathrm{hal} $ of the halo density, for $\Phi(r) \ll c^2$, is given by:
\begin{equation}
v^2_\mathrm{hal} (r) = \frac{G M(r)}{r} - \frac{1}{3} \Lambda c^2 r^2.
\label{eq:hal}
\end{equation}
For higher values of $\Lambda$, the maximum possible velocity decreases. It is possible to normalize the halo velocity term, Eq.~(\ref{eq:hal}), by using the overdensity~$\Delta_c$, as
\begin{equation}
\left( \frac{v_\mathrm{hal}(x)}{v_s}\right) ^2 = \frac{\bar{M}(x)}{x} - \frac{2}{3}\Delta_c^{-1} x^2,
\label{eq:halNorm}
\end{equation}
 {with $v_s := 2 r_s \sqrt{\pi G \rho_s}$ and the overdensity parameter $\Delta_c := \rho_s/\rho_\Lambda$.  The associated energy density of $\Lambda$ is}
\begin{equation}
\rho_\Lambda \equiv \frac{\Lambda c^2}{8 \pi G} = \left( 5.78 \pm 0.11 \right)\times 10^{-27}\, \mathrm{kg/m^2},
\end{equation}
 {where the value of the cosmological constant is taken from the Planck collaboration~\cite{Planck:2018vyg}. }

 {The fit for the flat rotation curves give an upper limit for the $\Delta_c^{-1}$ parameter and therefore an upper limit for $\rho_\Lambda$.  Figure~\ref{fig:NFWvel} shows schematically the flat-scaled rotation curves of a typical galaxy with the NFW profile of the dark matter density, without and with different values of the  {overdensity} parameter. The impact of $\Lambda$ reduces the velocity for large radii.}

\begin{figure*}[t!]
\centering
\includegraphics[width=0.23\textwidth]{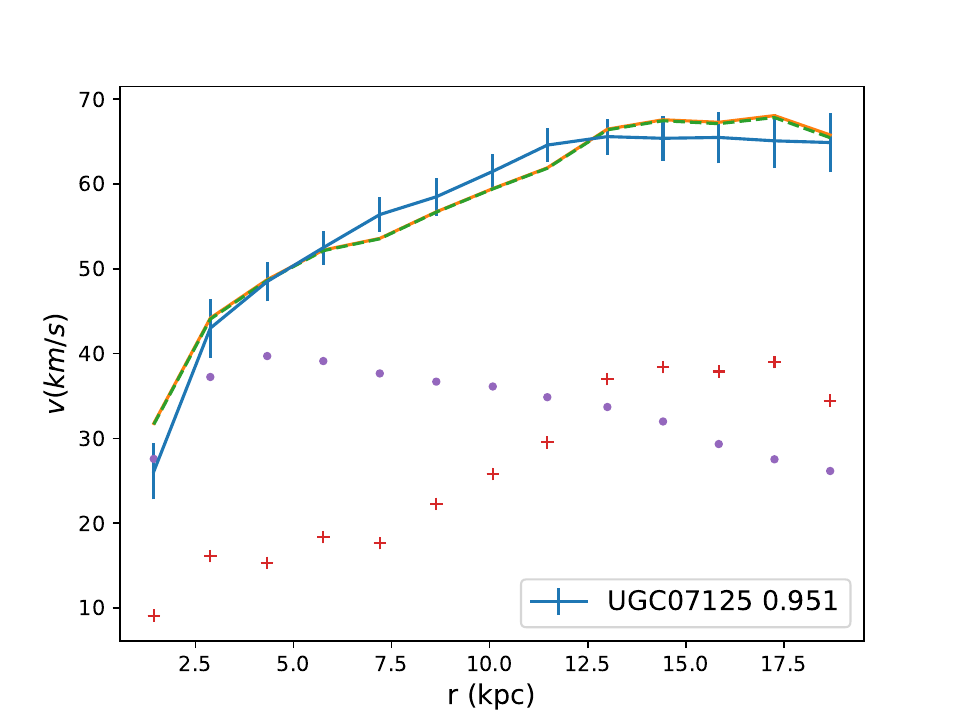}
\includegraphics[width=0.23\textwidth]{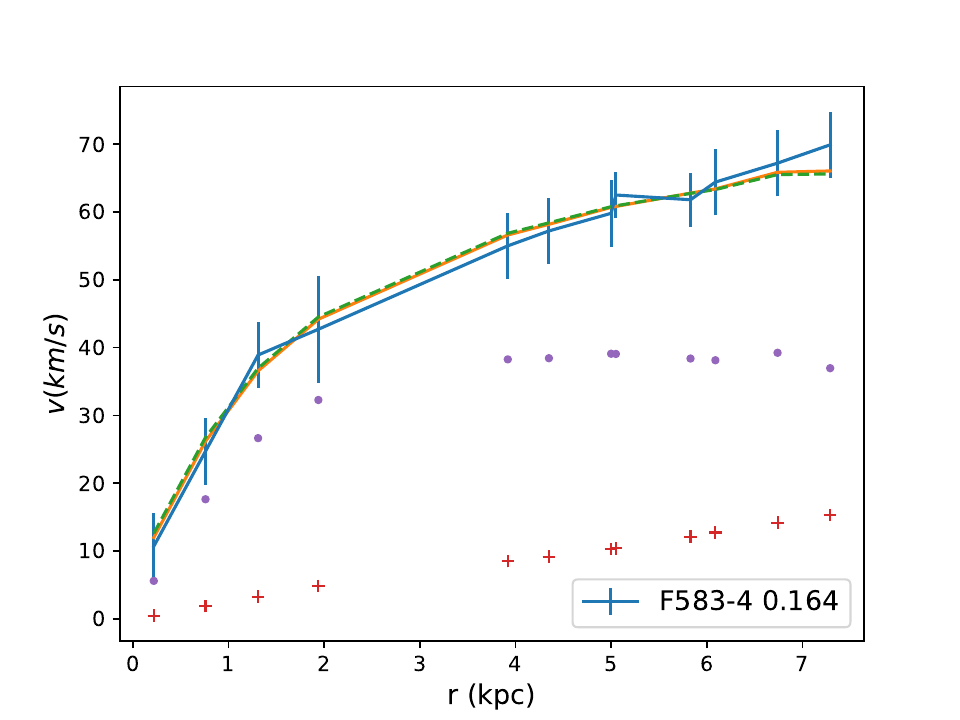}
\includegraphics[width=0.23\textwidth]{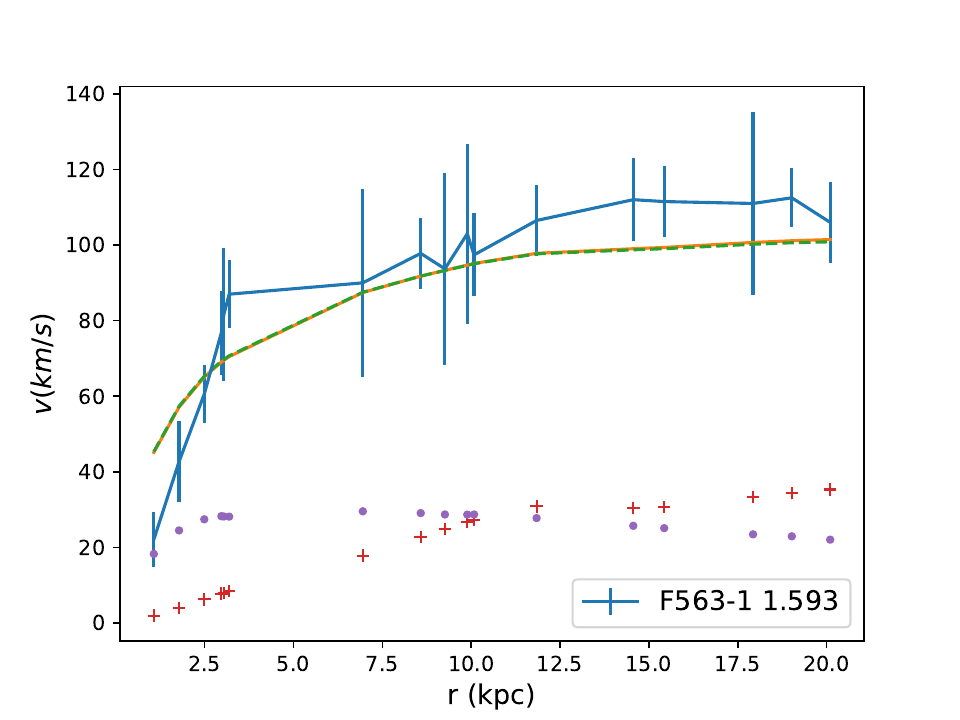}
\includegraphics[width=0.23\textwidth]{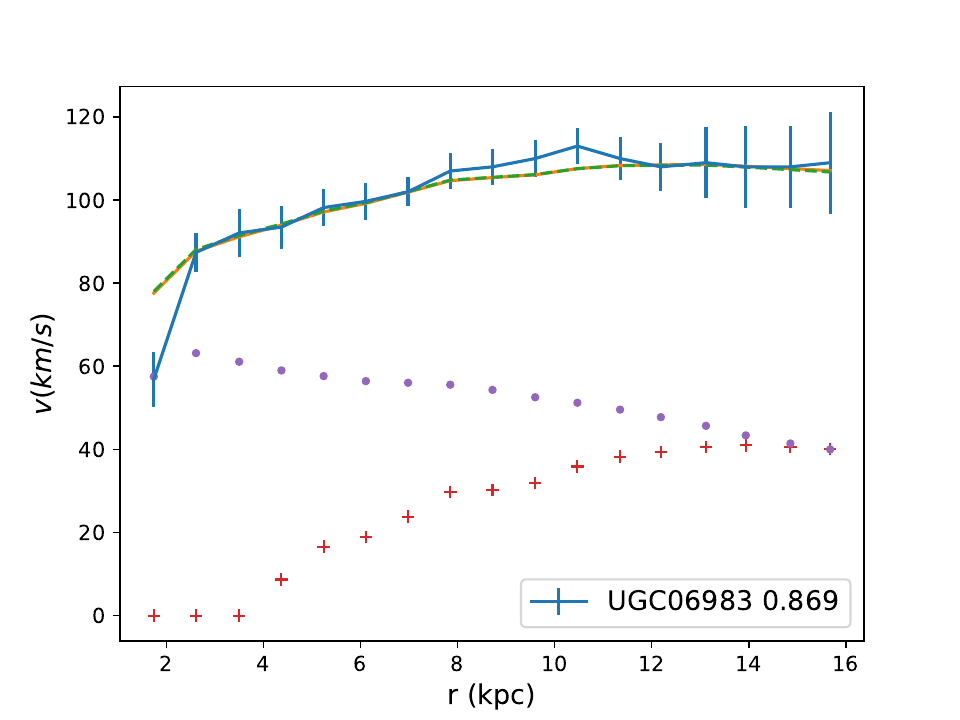}
\\
\includegraphics[width=0.23\textwidth]{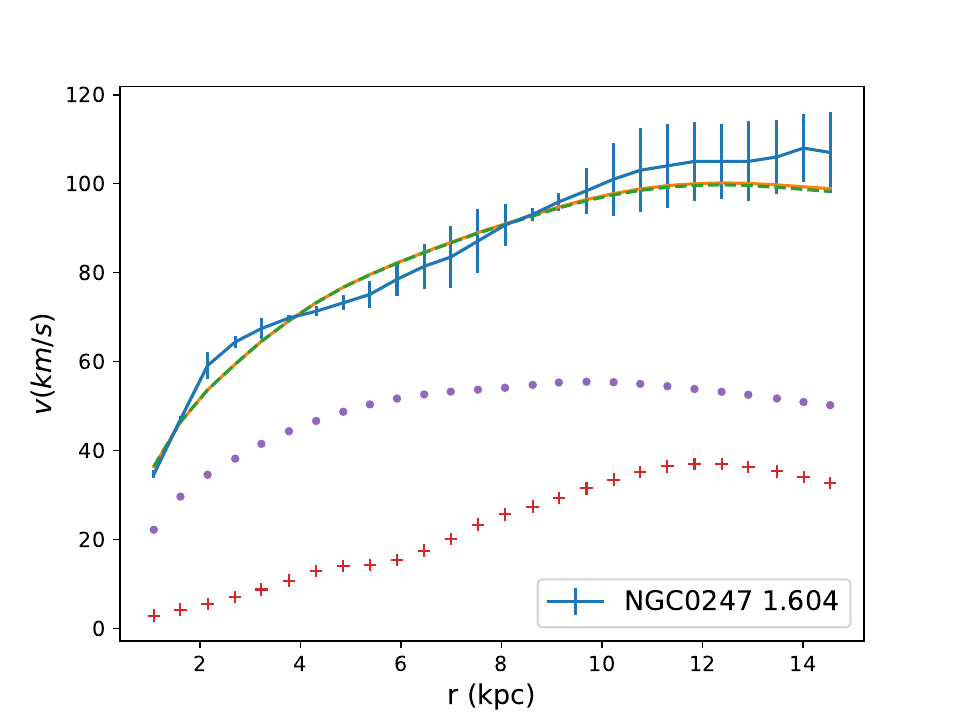}
\includegraphics[width=0.23\textwidth]{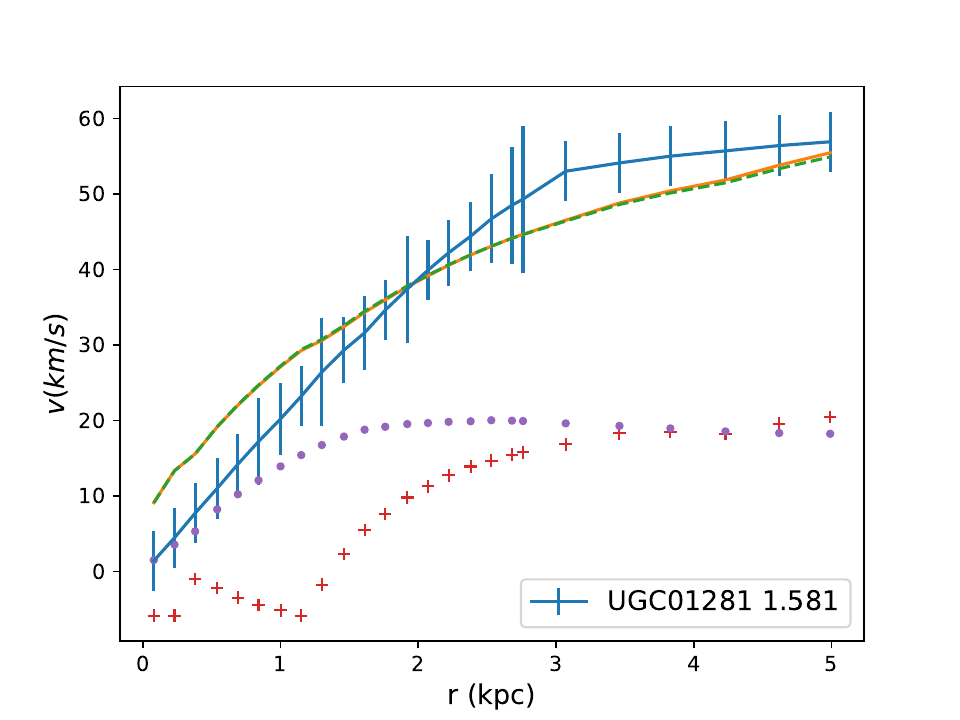}
\includegraphics[width=0.23\textwidth]{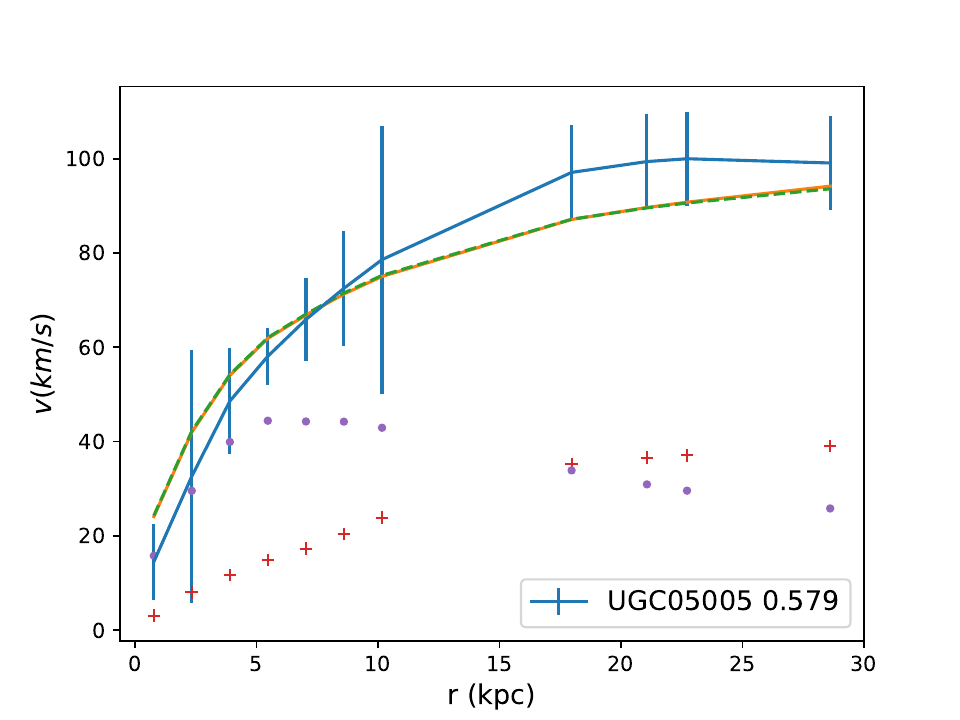}
\includegraphics[width=0.23\textwidth]{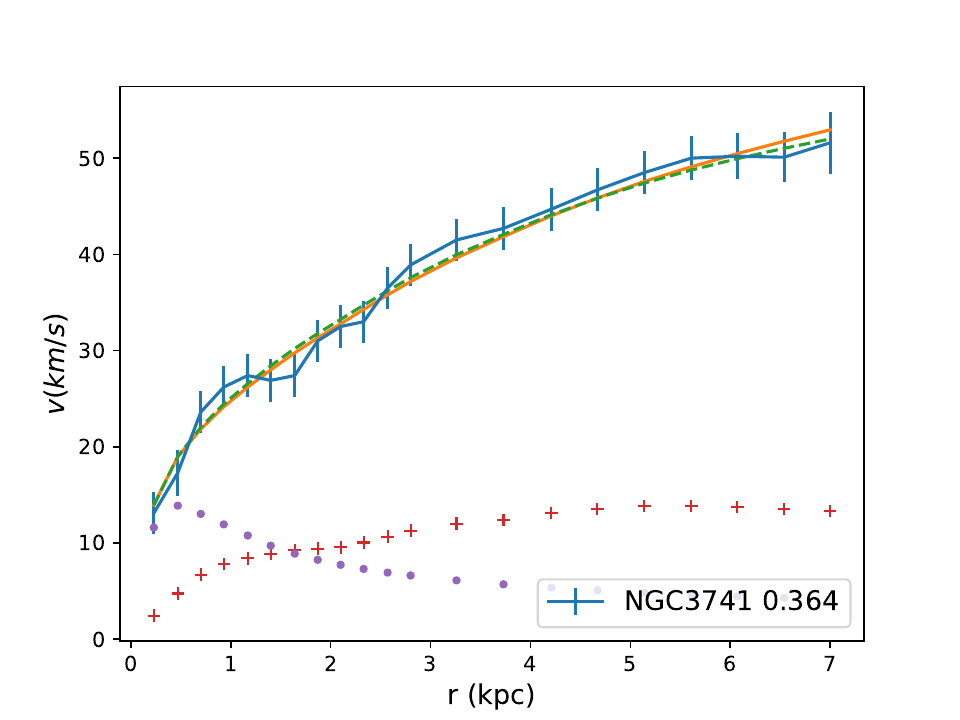}
\\
\includegraphics[width=0.23\textwidth]{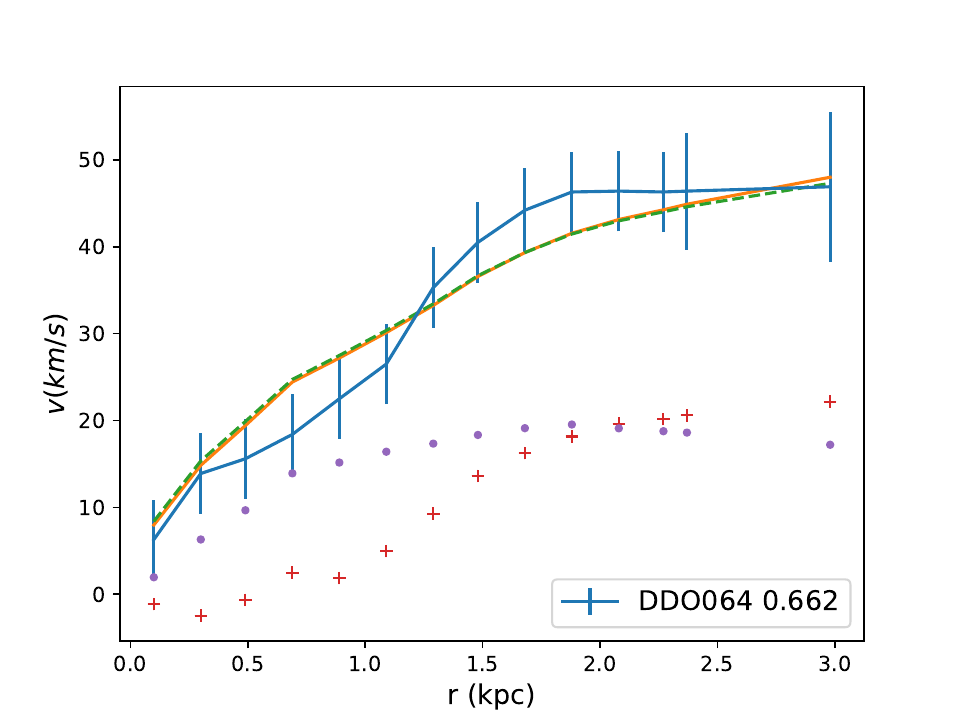}
\includegraphics[width=0.23\textwidth]{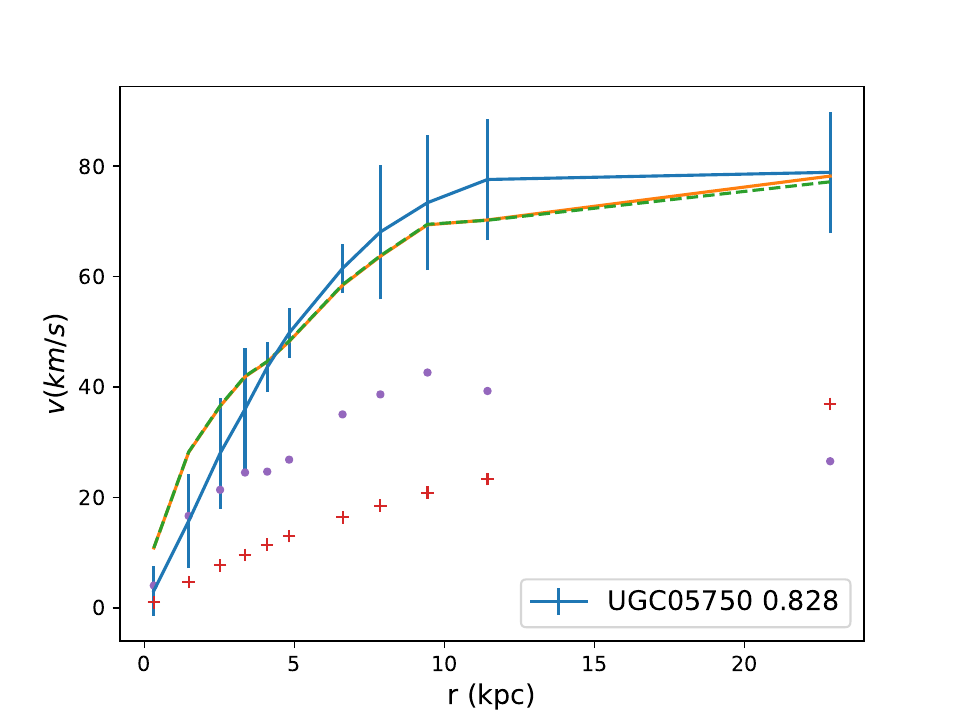}
\includegraphics[width=0.23\textwidth]{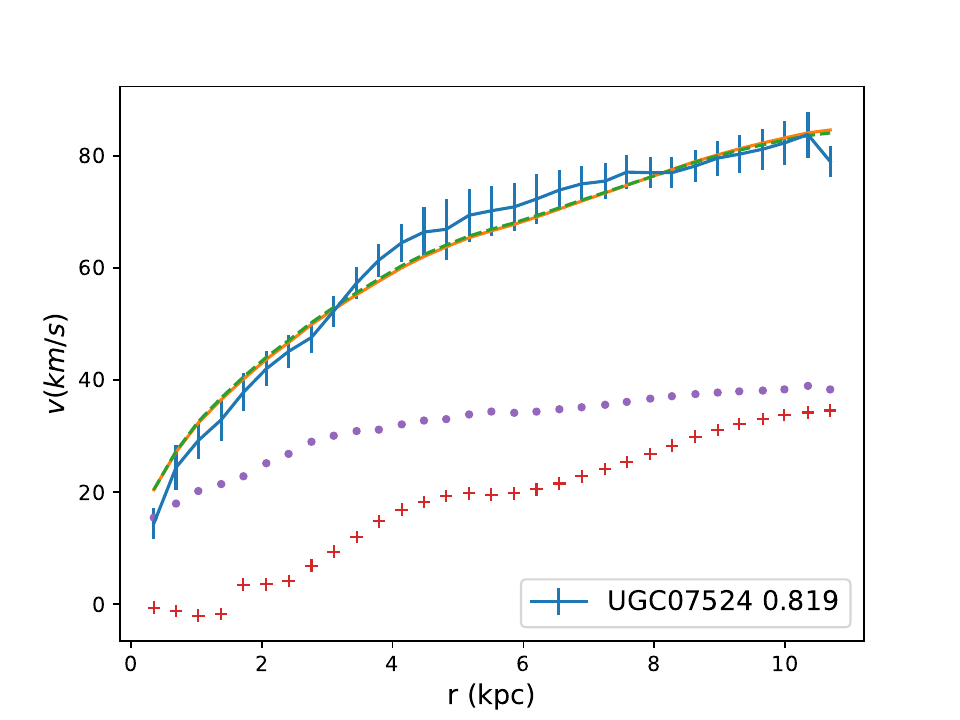}
\includegraphics[width=0.23\textwidth]{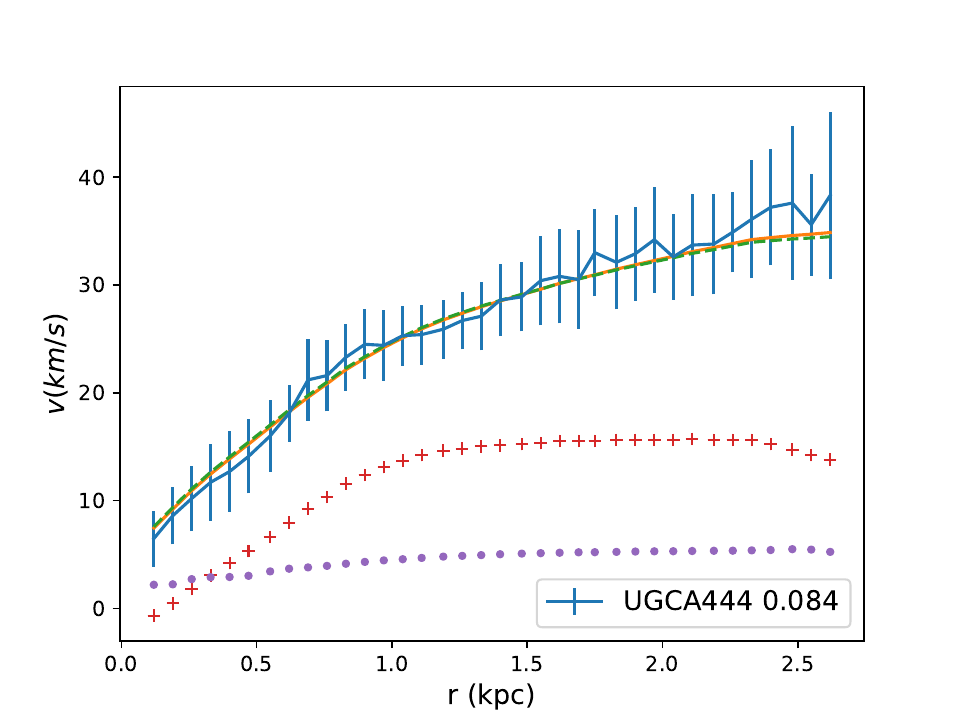}
\\
\includegraphics[width=0.23\textwidth]{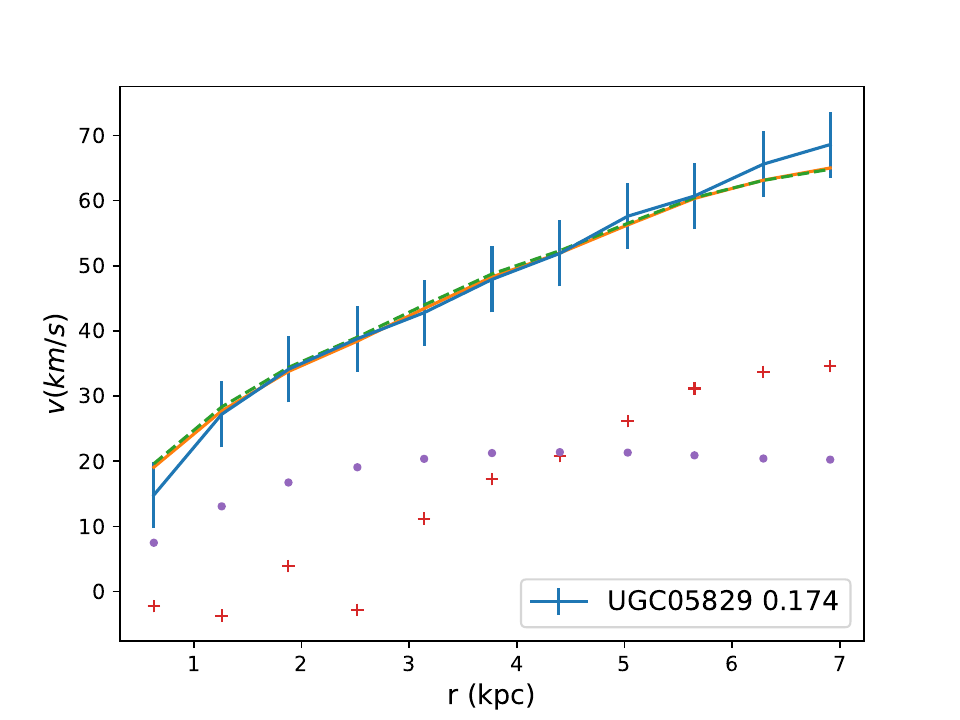}
\includegraphics[width=0.23\textwidth]{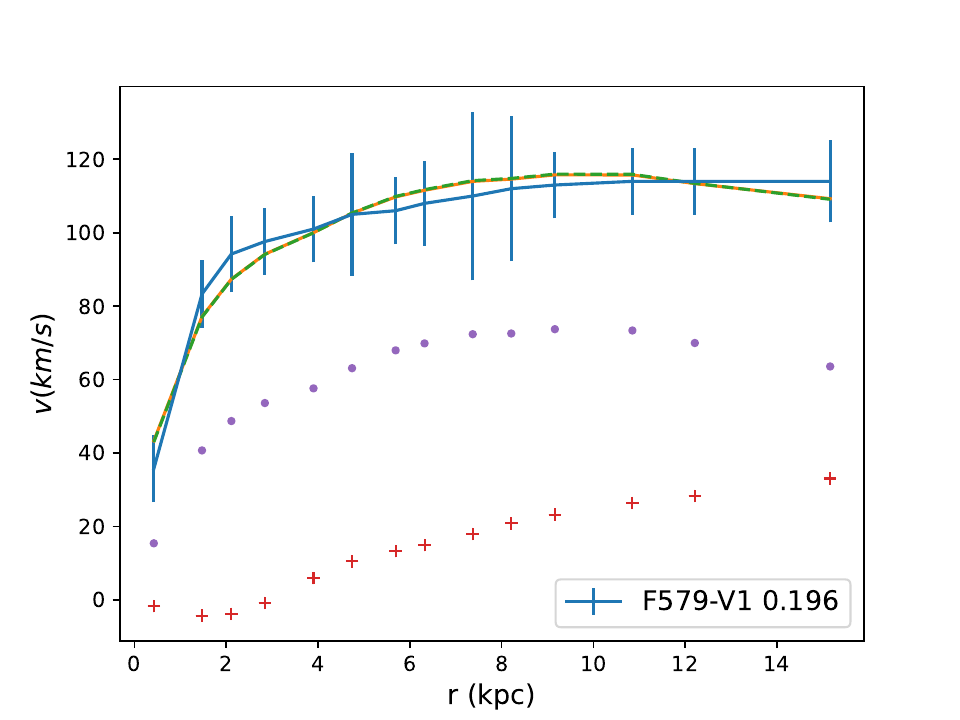}
\includegraphics[width=0.23\textwidth]{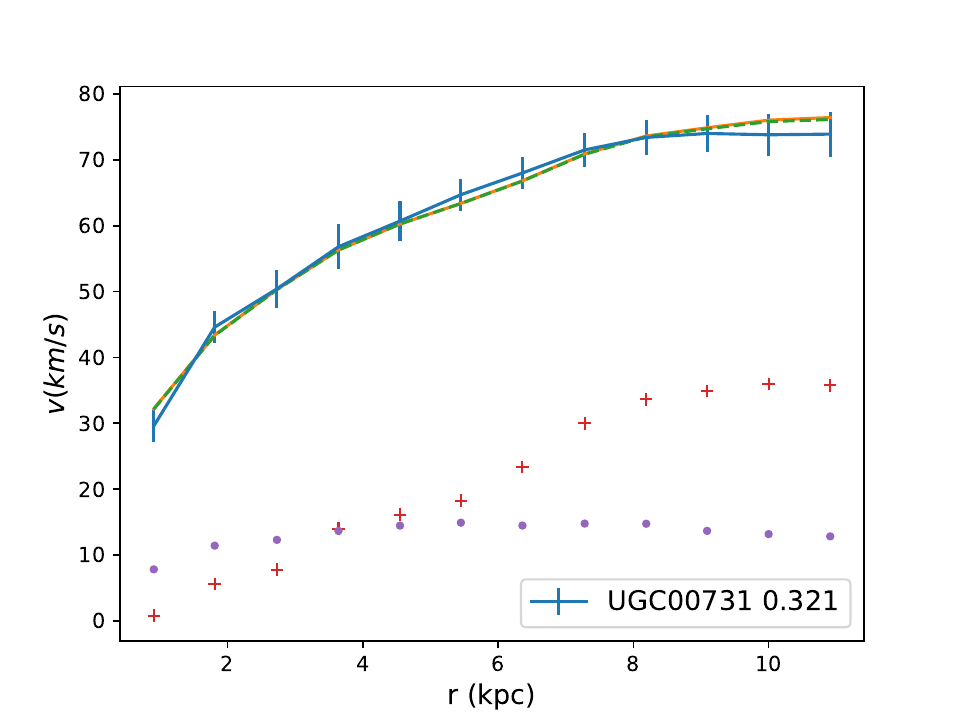}
\includegraphics[width=0.23\textwidth]{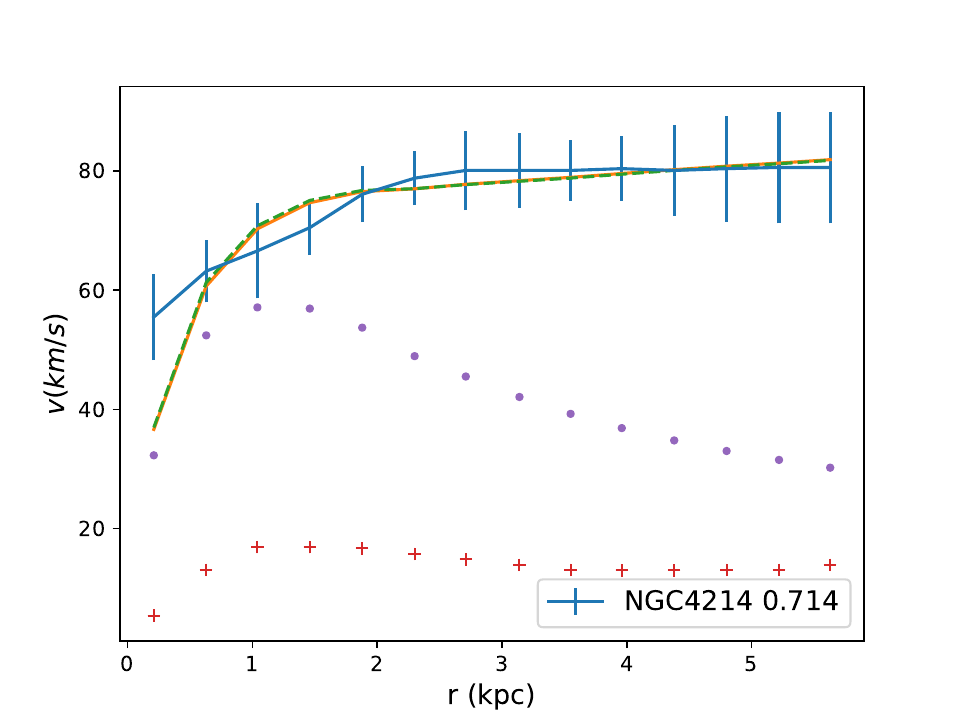}
\caption{\it{Examples for fits of different galaxies with the NFW profile without $\Lambda$ (smooth orange) for the NFW model.  {The observed velocity $v_{\text{obs}}$ with the corresponding errors is in blue, while the fit is in orange. The velocity of the gas is marked with ``+'', and the velocity of the disk with ``.''.} The reduced $\chi^2$ is also presented. In general, the models are sufficiently flexible to reproduce the observational data to good accuracy.}}
\label{fig:fits}  
\end{figure*}

\section{Bounds on \texorpdfstring{$\Lambda$}{Lambda}}
\label{sec:method}

\subsection{Analysis methodology}
\begin{figure*}
\centering
\includegraphics[width=0.9\textwidth]{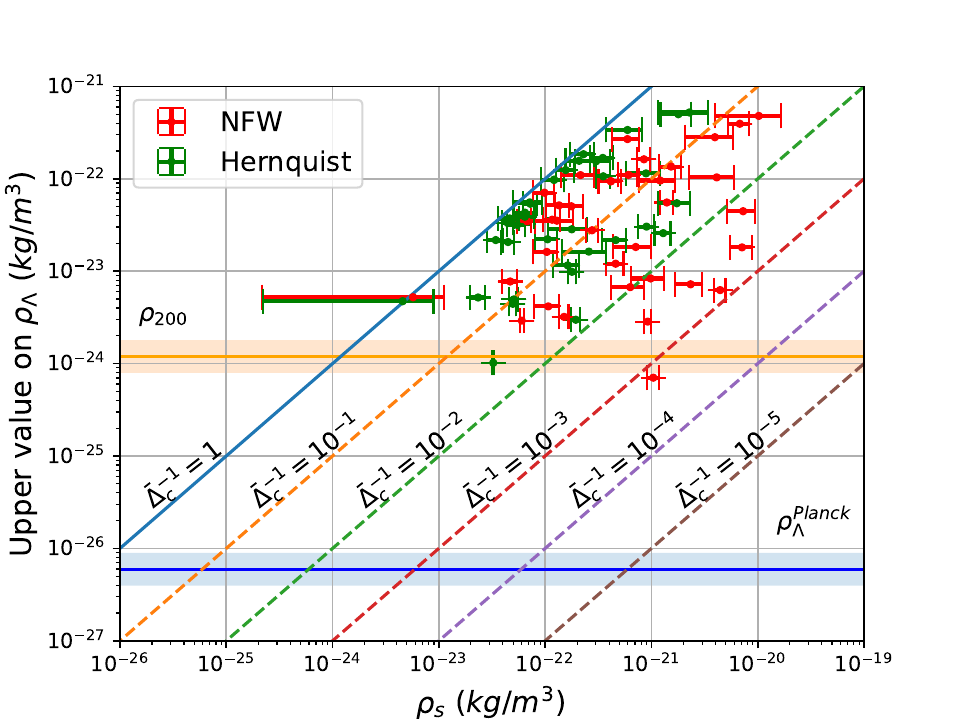} 
\caption{\it{The upper limits on $\Lambda$ vs. the corresponding scaling density for different galaxies for the NFW (red) and the Hernquist models (green) with a cosmological constant presence. The lines above corresponds to different values of  $\bar{\Delta}_c^{-1} = \rho_{(<\Lambda)}/\rho_s$. The area above the dashed blue line, $\bar{\Delta}_c^{-1} > 1$, is a ``forbidden area'' where the $\Lambda$ energy density is larger then the scaling density of the galaxy. The galaxies fit gives a wide range of $\bar{\Delta}_c^{-1}$ in between $1$ and $10^{-3}$. The upper best-fit limit (defined by the deviation $1\sigma$ for the mean posterior value) is around $\rho_{200}$.}}
\label{fig:posterior} 
\end{figure*}
The predicted circular velocity $v_\mathrm{pred}$ in the galactic plane is broken down into contributions from the gas, disk and dark matter halo:

\begin{equation}\label{eq:vpred}
v_\mathrm{pred}^2 =  v_\mathrm{hal}^2 + v_\mathrm{gas}^2+ \Upsilon_\mathrm{disk}\,v_\mathrm{disk}^2.
\end{equation}
We test the upper limit for $\Lambda$ with the SPARC dataset~\cite{Lelli:2016uea,McGaugh:2016leg,2016ApJ...816L..14L,2017ApJ...836..152L,2017MNRAS.466.1648K,Lelli:2017sul,Desmond:2018oai,Li:2018rnd,Li:2018tdo,Katz:2018wao,2018NatAs...2..924M,2018MNRAS.480.2292S,2018MNRAS.480.4287K,Li:2019zvm,McGaugh:2019auu,Lelli:2019igz,2019MNRAS.483.1496S,Street:2022nib}. The SPARC database includes $175$ late-type galaxies with high-quality rotation curves detected via near-infrared Spitzer photometry. The measurements allow tracing the rotation velocity out to large radii, providing strong constraints on the dark matter halo profiles. The mass models for the stellar disk are built by numerically solving the Poisson equation for the observed surface brightness profile. The derived gravitational potentials of the baryonic components are represented by the circular velocities of the test particles, tabulated as $v_\mathrm{disk}$, $v_\mathrm{bul}$, and $v_\mathrm{gas}$ corresponding to the contributions of the stellar disk, bulge, and gas, respectively.

\begin{figure}
     \centering
     \includegraphics[width=0.9\linewidth]{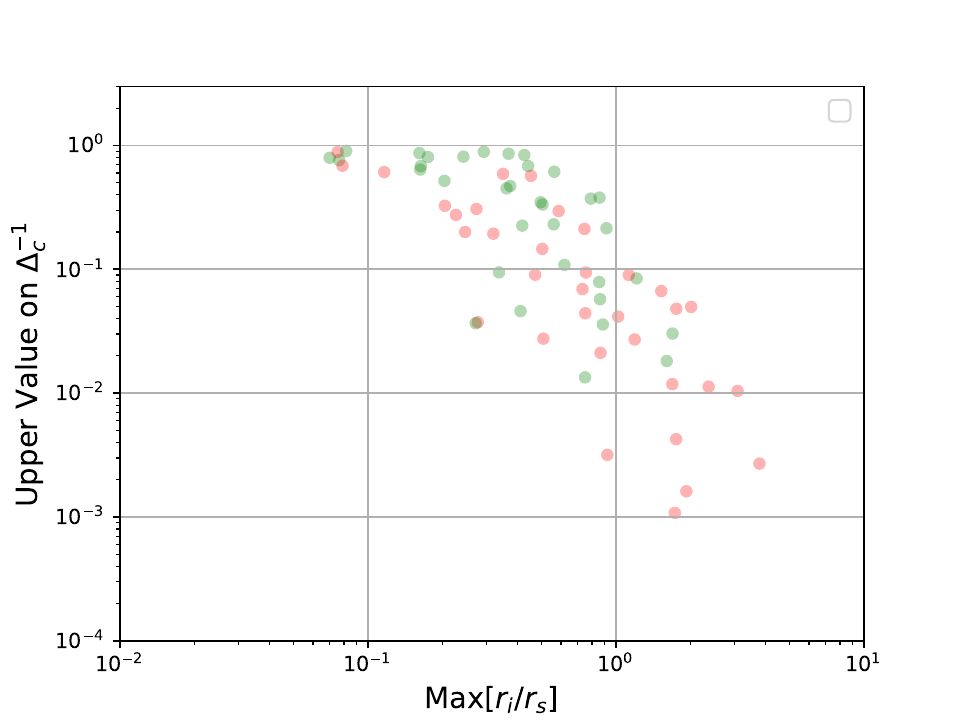}
\caption{\it{  {The correlation between the last normalized data point (over the fitted $r_s$) vs the upper limit on $\Lambda$. The correlation shows that an increase in sampling points in the outskirts leads to a decreased upper limit on $\rho_\Lambda$.}}}
\label{fig:corr} 
     \includegraphics[width=\linewidth]{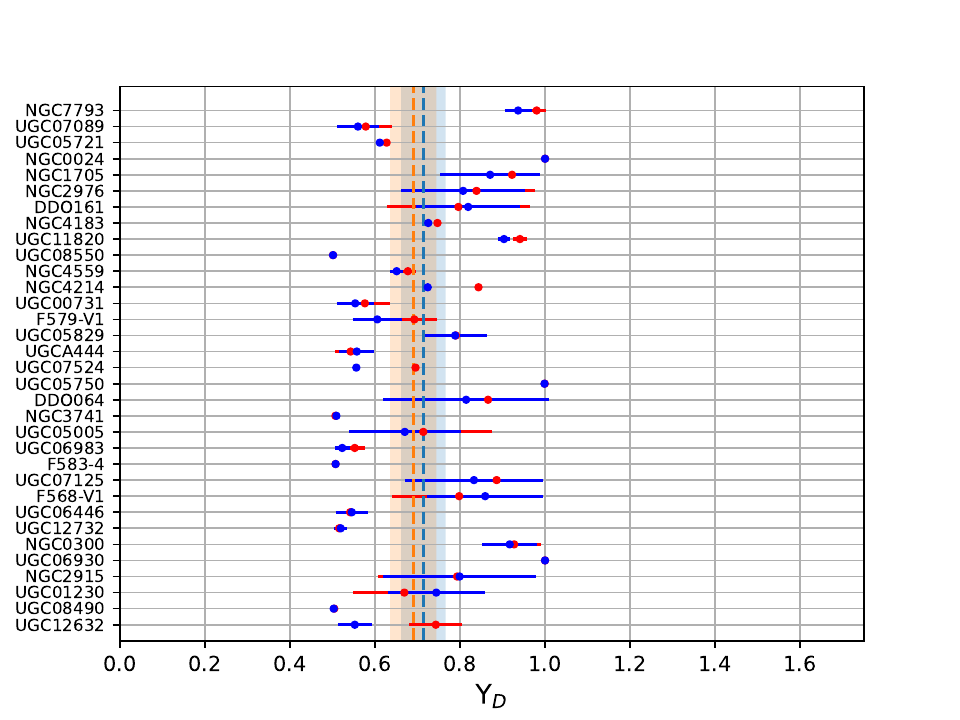}
\caption{\it{  {The final fitted mass to light ratio $\Upsilon_D$ for the good fit galaxies using the NFW density profile (blue) and the Hernquist density profile (red) with $1\sigma$. The average value of the mass-to-light ratio is $\bar{\Upsilon}_D = 0.79 \pm 0.11$ for the NFW model and $\bar{\Upsilon}_D = 0.74 \pm 0.10$ for the Hernquist model.}}}
\label{fig:mastolightratio}  
\end{figure}

Contributions $v_\mathrm{gas}$, $v_\mathrm{disk}$ (and $v_\mathrm{bul}$) are deduced from the HI gas density and light profiles, respectively, and provided by SPARC. The $\Upsilon_\mathrm{disk}$ is the stellar mass-to-light ratio for the disk (bulge) and it is equivalent to the mass $M_D$ of the disk divided by the luminosity $L_D$ of the disk. We choose galaxies without bulge contribution to reduce the degeneracy with other degrees of freedom to the fit.

We use Bayesian analysis to find the best parameters, $r_s, \rho_s$ implicitly through $v_s$, and $\Lambda$, for each galaxy vs. the observational data. We assume that the errors of the observed rotation curve data follow a Gaussian distribution, so that we can build the $\chi^2$ defined as:
\begin{equation}
\chi^2 = \sum_i\left(\frac{v_\mathrm{obs}^{(i)} - v_\mathrm{pred}(r_i,\rs,v_s, \Upsilon_\mathrm{disk},\bar{\Delta}_c^{-1} )}{\sigma_i}\right)^2
\end{equation}
 {where $v_\mathrm{obs}^{i}$ is the observed velocity with error $\sigma_i$, while the prediction is $v_\mathrm{pred}$, \eref{eq:vpred}. To give the upper value of $\Lambda$ as a repulsion force, we test the zero-point surface where $v = 0$ or $\Delta_c^{-1} = 1$. For the likelihood maximization, we use an affine-invariant MCMC nested sampler, as it is implemented within the open-source package \texttt{Polychord}~\cite{Handley:2015fda} with the \texttt{GetDist} package~\cite{Lewis:2019xzd}, to present the results.}

We use the following flat priors: $v_s \in [0, 200] \, \mathrm{km/sec}$, $r_s \in [0,50]\,\mathrm{Kpc}$, $\Upsilon_D \in [0.2, 1.5]$ and $\bar{\Delta}_c^{-1} \in [0,1]$. The upper limit of $\rho_\Lambda$ is noted as $\rho_{(<\Lambda)}$, and is associated with the  {overdensity} $\bar{\Delta}_c^{-1}$ via: $\bar{\Delta}_c^{-1} := \rho_{(<\Lambda)}/\rho_s$. We selected galaxies with more than 7 data points, which left us with 93 suitable galaxies. After fitting the models, we consider only ``good fit models'' satisfying
\begin{equation}
\chi_{\mathrm{red}}^2 = \chi^2/(N - d + 1) \leq 1.7,
\end{equation}
where $\chi_{\mathrm{red}}^2$ is the reduced $\chi^2$, normalized by the number of degrees of freedom, calculated as the number of data points for an individual galaxy $N$ minus the number of parameters $d = 4$ plus one. Using this second selection criterion, we identify around 50 galaxies suitable for our analysis.

\subsection{Results}
 {Fig.~(\ref{fig:fits}) shows examples of good fit for the NFW model. The observed velocity with the corresponding errors is colored blue, whereas the fit is colored orange. The velocity of the gas is marked with ``+'' and the disk velocity  with ``.''.  The reduced $\chi_{\text{red}}^2$ is also presented. The fit including $\Lambda$ (orange dashed lines) is similar to the fit for the purely Newtonian potential, coupled to the dark matter model.} This implies only a minor impact of $\Lambda$, and the fit yields an upper limit for $\Lambda$. We define this upper limit as the confidence bound $1\sigma$ around the mean. Figure~\ref{fig:posterior} shows the upper limit on $\Lambda$ vs.\ the corresponding scaling densities of different galaxies for the NFW (red) and the Hernquist models (green).

The lines correspond to $\bar{\Delta}_c^{-1} := \rho_{(<\Lambda)}/\rho_s$. The area above the dashed blue line, $\bar{\Delta}_c^{-1} > 1$, is the ``forbidden area'' where the repulsion force of $\Lambda$ is so strong that the galaxy would not exist anymore. The possibility of a virilized close galaxy from the fit exists with $1\sigma$ probability. Besides one outlier, a galaxy where the posterior is with a huge error bar, all galaxies obey this principle and have $\bar{\Delta}_c^{-1} \ll 1$ in the range $10^{-1}$ to $10^{-3}$. We define the upper limit on $\rho_\Lambda$ as the mean+$1\sigma$ value. Therefore, the upper limit of $\rho_{\Lambda}$ is in the range~$10^{-22}$~ kg/m$^3$ to $10^{-25}$~kg/m$^3$ for the lowest cases. In order to compare results from different galaxies, we also display the value of~$\rho_{\Lambda}$ by Planck together with the~$\rho_{200}$ value. According to the virialization of spherical density over cosmological background, $\rho_{200}$ is the corresponding value of the density at the outskirts of the galaxies.  {We expect that the upper limit on $\Lambda$ will be reduced with more data points in the galaxy outskirts.} Since the measurements to date are only around the~$r_s$ region, the upper limit on~$\Lambda $ approaches $\rho_{200}$ only for a subset of the sample. Figure~\ref{fig:posterior} shows that albeit the NFW model predicts slightly higher values, the results are robust for different models of dark matter. Since the NFW model is more cuspy in the center and the Hernquist shallower, the former model predicts higher values.

The range of the upper bound on $\Delta_c^{-1}$ 
is quite broad for different galaxies. To explain why the range changes for different galaxies, we consider the correlation between the maximal $r_i/r_s$ of a certain dataset vs. the upper limit on $\Delta_c^{-1}$. Figure~\ref{fig:NFWvel} shows that the impact of $\Lambda$ takes place on the outskirts of the galaxy. Figure~\ref{fig:corr} shows the correlation between these quantities.  {It is clear from the correlation that with more points in the outskirts of the galaxy ($r > r_s$) the upper limit of~$\rho_\Lambda$ is lower.}

 {To confirm the correlation between the upper limit on $\Delta_c^{-1}$ and the relative last distance point $r_i/ r_s$, quantifying how far the measurements cover the outskirts of the galaxy as shown in Fig.~(\ref{fig:corr}), we use the Spearman rank correlation~\cite{ca468a70-0be4-389a-b0b9-5dd1ff52b33f} which is a non-parametric measure of rank correlation for monotonic relations.} For a perfect monotonic behavior, the Spearman correlation is $+1$ for increasing behavior (and $-1$ for decreasing). We find the correlations $-0.85$ for the NFW profile, and $-0.69$ for the Hernquist profile which confirms our expectation. Fig. ~(\ref{fig:mastolightratio}) shows the final fitted values of $\Upsilon_D$  for the good-fit galaxies for the NFW density profile (blue) and for the Hernquist density profile (red). The average value of the mass to light ratio is~$\bar{\Upsilon}_D = 0.79 \pm 0.11$ for the NFW model, and~$\bar{\Upsilon}_D = 0.74 \pm 0.10$ for the Hernquist model. In the literature, $\bar{\Upsilon}$ reads $\sim 0.7$ which is compatible with our posterior distribution. 

\begin{figure*}[t!]
\centering
\includegraphics[width=0.75\textwidth]{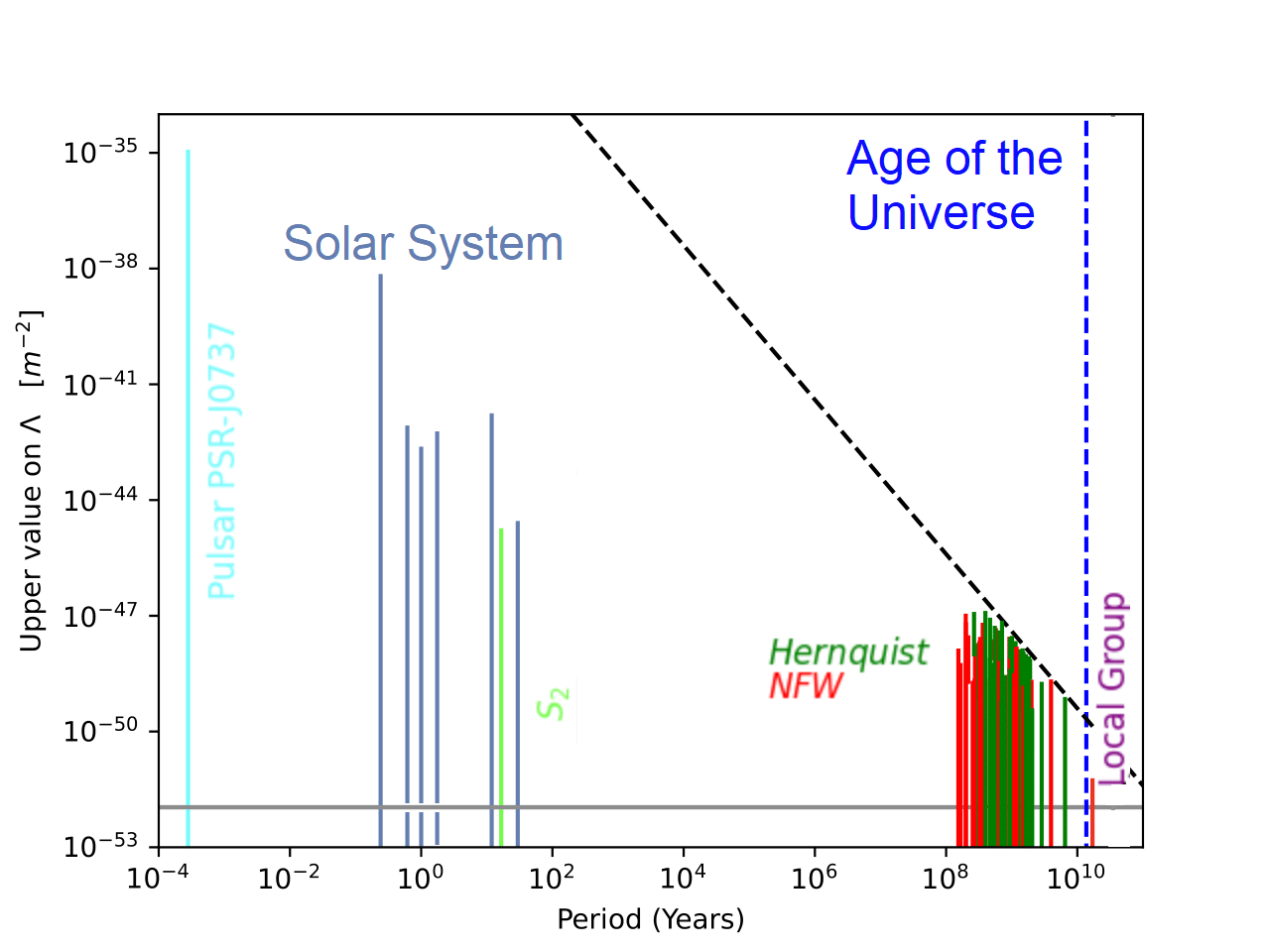} 
\caption{\it{Comparison between bounds on $\Lambda$ for different systems versus the curvature scalar for these systems: planets in the solar system (dark blue), S2 star around Sgr-A* (green), double pulsar PSR~J0737-3039A/B (azure), Local Group (purple) with the new fit discussed in~\cite{Benisty:2023clf}. Each upper limit is the $1\sigma$ upper limit on $\Lambda$. For comparison, we put the estimated values of $\Lambda$ from~\cite{Planck:2018vyg}. For longer periods in binary motions or lower spherical densities in galaxies, the upper limit on $\Lambda$ decreases and getting close to the one measured from Cosmology.  {The dashed black line represents $\Delta_c^{-1} = 1$ that separates bound (under the line) from unbound systems (above the line).}}}
\label{fig:com}  
\end{figure*}

We point out the limitation of analyzing galactic flat rotation curves, as advocated here, for deriving further information on DE models. The main obstacle to constraining DE models is the degeneracy between any additional parameters and~$\Lambda$. 
Take, as an example, Ref.~\cite{Zhang:2023neo} that generalizes the potential in Eq.~(\ref{eq:f}) by a variable Equation of State (EoS):
\begin{equation}
\phi = G M/r  + \left(6/(\Lambda r^2)\right)^{(3 w +1)/2}  , 
\end{equation}
where $w$ is the EoS parameter. For $w = -1$ the potential in Fig.~5 of Ref.~\cite{Zhang:2023neo} shows that the best fit of $w$ is degenerate with the best fit of the radius, $r_0 = \sqrt{6/\Lambda}$. Hence, constraining both parameters, $w$ and $r_0$, at once is not possible; only a relation between these two parameters can be derived. 

\section{Consistency with other probes}
\label{sec:probes}
 {The two conditions that were discussed for detecting $\Lambda$ are also compatible with other systems, even though they were based on different datasets. Ref.~\cite{Benisty:2024lsz} shows consistently that for binary systems (or effective one body systems) the upper limit on $\Lambda$ is reduced for systems with higher orbital frequencies and is possible to detect if the orbital frequency is about  $T_{\Lambda} = 2\pi/c \sqrt{3/\Lambda}\approx 35 \, Gy$. The equilibrium condition in Eq.~(\ref{eq:force}) gives: $GM/r^3 = \Lambda c^2/3$. The left-hand side is the orbital frequency of a binary system $\omega_{Kep } = \sqrt{G M/r^3}$ with total mass $M$ and separation $r$, and the right-hand side is the frequency $\Omega_\Lambda = 2 \pi/T_{\Lambda} = \sqrt{\Lambda c^2/3}$. This shows that $T_{\Lambda}$ is an important scaling or a critical period, where the Newtonian force is canceled by the repulsion force of $\Lambda$. Moreover,~\cite{Benisty:2023clf} finds a strong upper limit on $\Lambda$ when the period of the system is closer to $T_{\Lambda} $.}

The binary systems addressed in Ref.~\cite{Benisty:2023clf} include the Solar System planets, the S2 star around the galactic center, the Double Pulsar PSR-J0737, and the binary motion of the Milky Way and M31 galaxies. Considering measurements from different binary systems that have orbital periods from days to gigayears, the upper bound on the cosmological constant is shown to decrease as the orbital period of the system increases. Ref.~\cite{Benisty:2023clf} derives the upper limit on $\Lambda$ from the precession of solar system planets, the precession of the S2 star around the galactic center~\cite{GRAVITY:2020gka}, and the precession of the double pulsar PSR J0737-3039A/B~\cite{Kramer:2021jcw}, since the precession term is modified in the presence of $\Lambda$. Despite its more precise measurements, the upper bound on $\Lambda$ from the double pulsar is higher than the one obtained from S2 or Saturn, which yields the lowest constraint among these systems.

 {The case of rotating spherical density is similar. With the scaling density $\rho_s$ and the scaling period $T_s$ (at $r_s$), the frequency of the spherical orbit can be expressed as~\cite{1988gady.book.....B}:}
\begin{equation}
\omega_s = 2 \pi/T_s = \sqrt{4 \pi G \rho_s/3}.
\end{equation}
 {where the orbital frequency/period of the system $\omega_{\text{Kep}}$ is related to the curvature. Fig.~(\ref{fig:com}) plots the upper limits in $\Lambda$, derived from various astrophysical systems, vs. the corresponding curvature around those systems. Some of those systems are binary systems, while the systems discussed here are spherical densities. The same principle is also applicable here: Despite the double pulsar precession is much more accurate than the galaxies measurements (8 digits vs 2 digits of accuracy), the upper limit on $\Lambda$ from galaxies is much lower, since the period of the galaxies is closer to $T_\Lambda$. }

\section{Summary and Discussion}
\label{sec:dis}
 {This work applies the observational data from the SPARC dataset to constrain the Cosmological Constant ($\Lambda$) in spherically symmetric dark-matter halos.} We use the Navarro-Frenk-White (NFW) and Hernquist models to identify the most suitable galaxies for these models. $\Lambda$ acts as a repulsive force on galactic dark-matter halos agglomerated via Newtonian attraction. We show that albeit the $\Lambda$ scale is related to a length scale, the upper limit on $\Lambda$ is correlated to the density of the surroundings and the efficiency of the measurements. The fits of the  {overdensity}, $\bar{\Delta}_c^{-1}$, for selected galaxies give a wide range of values between $1$~and~$10^{-3}$. The fitted upper limit (defined by the deviation $1\sigma$ for the mean posterior value) sits around $\rho_{200}$ giving an upper limit on $\Lambda$ only two orders of magnitude larger than that measured in cosmological studies by the Planck Collaboration for the best cases.

To reduce the upper limit on $\Lambda$, we find two conditions: (i) the suitable environment (the curvature/period) and (ii) the quality of the data.  {The curvature of the environment is modeled by the Kretschmann scalar, which is proportional to the density of the halo. The density approaches low values on the outskirts of galaxies.} Therefore, the density is closer to $\rho_\Lambda$ for large radii and is expected to reduce the upper limit on $\Lambda$. We show that for the SPARC dataset, where the last point is $r_i \gg r_s$, the upper limit on $\Lambda$ is reduced. The second criterion is the quality of the data, which refers to good galaxy fitting explained by a dark matter model, bulge, and disk. For such known models, the upper limit on $\Lambda$ over these effects will also be reduced.

To improve the upper limit, future measurements must focus on the outskirts of galaxies, where the dark matter densities are closer to $\rho_\Lambda$ and the effect of $\Lambda$ is more pronounced. The \textbf{James Webb Space Telescope} aims to measure the outskirts of the Milky Way and other galaxies. In that case, we claim that the upper limit of $\Lambda$ could be reduced.

 {The method introduced in this paper has its limits, though:} (i) it relies on an empirical ad hoc model of the dark-matter halo, (ii) which is not necessarily spherical in nature, and (iii) the virial theorem only applies asymptotically. However, with more data on the radial velocities in the outskirts of the galaxies, we expect to improve the upper limit.

\acknowledgments
 {We thank the anonymous referee, Jenny Wagner, Paolo Salucci, and Anne-Christine Davis for useful discussions and comments.} DB and DV thank the Carl Wilhelm Fueck-Stiftung for support. DB is also grateful for the support from the Margarethe und Herbert Puschmann Stiftung and the European COST action CA21136.

\bibliographystyle{apsrev4-1}
\bibliography{ref.bib}

\end{document}